\documentclass[%
reprint,
superscriptaddress,
amssymb,
aps,
prl,
floatfix,
noeprint,
nonote
]{revtex4-2}

\usepackage{graphicx}
\usepackage{rotating}
\usepackage{amsmath}
\usepackage{bbm}
\usepackage{subfigure}
\usepackage{color}
\usepackage{braket}
\usepackage{tikz}
\usepackage{hyperref}
\hypersetup{colorlinks, 
	linkcolor={blue!75!black!80!yellow},
	citecolor={blue!75!black!80!yellow}, 
	urlcolor={blue!75!black!80!yellow}
	}

\makeatletter \renewcommand\@make@capt@title[2]{%
\@ifx@empty\float@link{\@firstofone}{\expandafter\href\expandafter{\float@link}}%
\sffamily{\textbf{#1}}\@caption@fignum@sep#2 }

\begin{document}

\title{Super-Poissonian Squeezed Light in the Ground State of Strongly Coupled Light-matter Systems} 

\author{Cankut Tasci}
\affiliation{Department of Physics, City College of New York, New York, NY 10031, USA}
\affiliation{Department of Physics, The Graduate Center, City University of New York, New York, NY 10016, USA}
\author{Mohammad Hassan}
\affiliation{Department of Physics, City College of New York, New York, NY 10031, USA}
\affiliation{Department of Physics, The Graduate Center, City University of New York, New York, NY 10016, USA}
\author{Leon Orlov-Sullivan}
\affiliation{Department of Physics, City College of New York, New York, NY 10031, USA}
\author{Leonardo A. Cunha}
\affiliation{Center for Computational Quantum Physics, Flatiron Institute, New York, 10010  NY,  USA}
\author{Johannes Flick}
  \email[Electronic address:\;]{jflick@ccny.cuny.org}
\affiliation{Department of Physics, City College of New York, New York, NY 10031, USA}
\affiliation{Department of Physics, The Graduate Center, City University of New York, New York, NY 10016, USA}
\affiliation{Center for Computational Quantum Physics, Flatiron Institute, New York, 10010  NY,  USA}
\date{\today}


\date{\today}

\begin{abstract}
Strong light-matter coupling enables hybrid states in which photonic and electronic degrees of freedom become correlated even in the ground state. While  many-body effects in long-range dispersion interactions are known to reshape electronic properties under such conditions, their impact on quantum-optical observables remains largely unexplored. Here, we address this problem using quantum electrodynamical density-functional theory (QEDFT) combined with the recently developed photon-many-body dispersion (pMBD) functional, which can capture higher-order electron-photon correlations and multi-photon processes. 
We compute ground-state photonic observables including photon number fluctuations, second-order correlations, and quadrature variances, and find  squeezing and super-Poissonian photon statistics emerging from light-matter interactions in the strong coupling regime. Our results demonstrate that capturing the full hierarchy of many-body, electron-photon and multi-photon correlations is essential for a consistent description of quantum-optical properties in strongly coupled molecular systems, establishing QEDFT as a first-principles framework for predicting nonclassical photonic features in the ground state of complex systems.

\end{abstract}

\maketitle

\textit{Introduction}: The emergence of strong light–matter interactions has opened new frontiers in chemistry and materials science~\cite{vidal2021manipulating}. In the strong coupling regime, when molecular or solid-state systems are embedded in optical or plasmonic cavities, matter excitations can hybridize with quantized electromagnetic modes, leading to novel chemical and physical properties. Particularly, the collective strong-coupling regime, where many molecules couple simultaneously to a cavity mode, has been the focus of numerous experimental efforts~\cite{thomas2019tilting,bxiang2020i, ahn2023modification, patrahau2024}.
Recent studies~\cite{ haugland2023understanding,philbin2023molecular,Cao_2025,tasci2025pmbd} have demonstrated cavity-induced modifications of scaling relations in long-range dispersion and van der Waals interactions. In this regime, higher-order effects of the light-matter interaction can be expected to become important, as many-body effects are known to be essential already for electronic dispersion interactions~\cite{donchev2006}.

A complementary perspective on strongly coupled light-matter systems arises from quantum optics and cavity QED, where the statistical properties of the field serve as sensitive probes of the underlying light-matter interaction~\cite{Carmichael1985, Rempe1991}.
Quantities such as the photon-number distribution ($\langle \hat{a}^\dagger \hat{a} \rangle$), second-order correlations ($\langle \hat{a}^\dagger \hat{a}^\dagger \hat{a} \hat{a} \rangle$), and quadrature variances ($\langle \hat{q}^2 \rangle, \langle \hat{p}^2 \rangle$) capture the emergence of collective effects and non-classical coherence.
These observables become relevant in phenomena such as superfluorescence and superradiance~\cite{DeVoe1996, Brandes2005}, photon antibunching in single-emitter systems~\cite{Basche1992, Michler2000}, and the generation of squeezed vacuum states~\cite{Breitenbach1997, Polzik1992}.
To understand how macroscopic collective coupling modifies these signatures in molecular systems, a balanced treatment of quantum electrodynamics and electronic structure is necessary.

One approach for studying strongly coupled light-matter systems is quantum electrodynamical density-functional theory (QEDFT). QEDFT provides an \textit{ab initio} framework for coupled electron-photon systems, treating electronic and photonic degrees of freedom on equal footing~\cite{tokatly2013,ruggenthaler2014}. However, most applications of QEDFT to date have focused primarily on studying changes in the electronic structure, while the explicit treatment of quantum-optical observables, in particular collective coupling and many-photon effects, remains limited. This past focus can be primarily explained by limitations of existing exchange-correlation (xc) functionals currently available within QEDFT \cite{pellegrini2015,flick2018a,flick2022,schafer2021making,lu2024electronphoton,mejia2025meta}. Many approximations neglect higher-order and multi-photon excitations, which are expected to be important in the collective strong-coupling regime. The absence of xc functionals capable of capturing these effects has restricted the predictive power of QEDFT, particularly for quantum-optical signatures. 

In this work, we present an application of the photon many-body dispersion (pMBD) functional~\cite{tasci2025pmbd} to overcome these limitations and analyze ground-state quantum-optical observables for general light-matter coupled systems. The pMBD functional extends the original MBD framework ~\cite{tkatchenko2009, tkatchenko2012, tkatchenko2013} to systems under strong light-matter coupling and incorporates long-range many-body correlation effects between matter and quantized electromagnetic fields, thus enabling an accurate description of higher-order electron-photon processes that naturally arise in the collective regime. Because pMBD is explicitly many-body in both the electronic and photonic sectors, it allows these higher-order correlations to manifest directly in quantum-optical observables. Moreover, the pMBD formalism is constructed in a way that provides straightforward access to quantized photonic operators \cite{gori2023second}, including $\hat{a}$ and $\hat{a}^\dagger$, making possible the evaluation of quantum-optical quantities and multi-photon correlation functions within an \emph{ab initio} framework. 

\textit{Theory:} Due to the gauge freedom in quantum electrodynamics, the light-matter Hamiltonian may be expressed in different but physically equivalent representations. While the original pMBD Hamiltonian was formulated in the length gauge~\cite{tasci2025pmbd}, in this work we reformulate the pMBD formulism in the velocity gauge to simplify the calculation of photonic observables. In the velocity gauge, the operators $\hat{a}_\alpha$ and $\hat{a}^\dagger_\alpha$ directly correspond to degrees of freedom of the cavity mode~\cite{schaefer2020relevance}. We refer to the SI for the transformation from length to velocity gauge.

In the velocity gauge, the pMBD Hamiltonian for a non-relativistic system of $N_a$ atoms, and $N_p$ quantized photon modes under dipole approximation then reads as

\begin{align}
\label{eq:ham}
\hat{H}
= & \;
\frac{1}{2} \sum_{i=1}^{N_a} \sum_{a=1}^{3} 
\left( 
    i \hat{\nabla}_{ia} + \sum_{\alpha=1}^{N_p} \lambda_{\alpha a} \hat{q}_\alpha \sqrt{\alpha_{ia}} \omega_{ia}
\right)^2
+ \omega_{ia}^2 \chi_{ia}^2 \nonumber \\
& + \frac{1}{2} \sum_{i,j=1}^{N_a} \sum_{a,b=1}^{3}
\omega_{ia} \omega_{jb} \sqrt{\alpha_{ia} \alpha_{jb}} \chi_{ia} \, T^{ab}_{\text{LR},ij} \, \chi_{jb} \nonumber \\
& + \frac{1}{2} \sum_{\alpha=1}^{N_p} 
 \hat{p}_\alpha^2 + \omega_\alpha^2 \hat{q}_\alpha^2 ,
\end{align}

where $\chi_{ia} = \sqrt{m_i} \xi_{ia}$ describe the mass-weighted atomic displacements
with effective atomic frequencies $\omega_{ia}$, charges $e_i$, and
polarizabilities $\alpha_{ia} = e_i^2/(m_i \omega_{ia}^2)$~\footnote{We note that we follow regular procedure to obtain $\alpha_{ia}$ and related parameters via Hirshfeld decomposition and rsscs equations \cite{tkatchenko2012,tkatchenko2013}}. $T^{ab}_{\text{LR},ij}$ is  interaction tensor that describes the long-range dipole–dipole interaction between the individual atoms~\cite{ambrosetti2014long}. In the velocity gauge, the light-matter interaction is introduced via the coupling of the atomic momentum $i\hat \nabla_{ia}$ and the product of charge $e_i$ and vector potential $\lambda_{\alpha a} \hat{q}_\alpha$.

The velocity-gauge Hamiltonian in Eq.~\ref{eq:ham} contains now coupling between position and momentum degrees of freedom, requiring the diagonalization of a $2 (3N_a + N_p) \times 2 (3N_a + N_p)$ matrix. We follow the procedure of Ref.~\citenum{COLPA1978327}, which correctly accounts for the commutation relations between momentum and position operators (see SI for details). This scaling is slightly different from the length-gauge formulation,  where the position and momentum degrees of freedom formally become decoupled. As a consequence, the eigenvalues for the length-gauge pMBD Hamiltonian can be obtained by diagonalizing a $(3N_a + N_p) \times (3N_a + N_p)$ matrix. {We note that instead of diagonalizing the pMBD Hamiltonian in Eq.~\ref{eq:ham} directly, the pMBD energy can be equivalently obtained as an integral over the imaginary frequency, which allows for a many-body order-by-order expansion of the energy.

Beyond providing the total exchange-correlation energy~\cite{tasci2025pmbd}, the pMBD framework can be developed to access to a wide range of quantum-optical observables. 

\textit{Second Quantization:}  
To compute quantum-optical observables, it is both natural and advantageous to reformulate the pMBD Hamiltonian in a second-quantized framework. This formalism provides a transparent and compact way to describe collective excitations, enabling the use of operator algebra to directly access photonic observables. Here, we follow the procedure outlined in Ref.~\citenum{gori2023second} and, adapt it to the velocity gauge by following Ref.~\citenum{COLPA1978327}. We begin with the pMBD Hamiltonian in Eq.~\ref{eq:ham} expressed in terms of mass-weighted coordinates $(\chi_{ia}, q_\alpha)$ and their corresponding momenta. A bosonic representation is then introduced via canonical quantization, where all atomic and photonic degrees of freedom are mapped to ladder operators. In a first step in Eq.~\ref{eq:ham}, all position $\chi_{ia}$ and $q_\alpha$ operators and their conjugated momenta are replaced by (bosonic) creation and annihilation operators. In the next step, these original (bare) ladder operators are connected to the diagonalized collective normal modes $ \hat{b}, \hat{b}^\dagger $ through a Bogoliubov transformation:
\begin{align*}
    \begin{pmatrix}
        \hat{b} \\ \hat{b}^\dagger
    \end{pmatrix}
    =
    {\begin{pmatrix}
        X & Y \\
        Y & X
    \end{pmatrix}}
    \begin{pmatrix}
        \hat{a} \\ \hat{a}^\dagger
    \end{pmatrix}.
\end{align*}
The transformation matrices \( X \) and \( Y \) capture the hybridization between photonic and electronic modes, providing direct insight into vacuum fluctuations and correlation effects. The precise definitions of the \( X \) and \( Y \) matrices, along with detailed derivations and proofs of the transformation formulas, are provided in the SI. 

Having developed this connection now allows us to formulate various quantum-mechanical observables directly within the pMBD framework.
For a given photon mode $\alpha$ with frequency $\omega_{\alpha}$, the quadrature variances $\langle \hat{q}^2_{\alpha} \rangle$ and $\langle \hat{p}^2_{\alpha} \rangle$ in terms of the Bogoliubov matrices $X$ and $Y$ are given by 
\begin{align*}
\langle \hat{q}^2_{\alpha} \rangle
&= \frac{1}{\omega_{\alpha}}
\left[
\big( Y^{\mathrm T} Y \big)_{\alpha,\alpha}
-
\big( X^{\mathrm T} Y \big)_{\alpha,\alpha}
+
\frac{1}{2}
\right], \\[4pt]
\langle \hat{p}^2_{\alpha} \rangle
&= \omega_{\alpha}
\left[
\big( Y^{\mathrm T} Y \big)_{\alpha,\alpha}
+
\big( X^{\mathrm T} Y \big)_{\alpha,\alpha}
+
\frac{1}{2}
\right].
\end{align*}
Because the pMBD Hamiltonian in Eq.~\ref{eq:ham} is a quadratic Hamiltonian, its ground state is a zero-mean Gaussian state~\cite{weedbrook2012} where all first moments of the position and momentum operators (i.e. $\langle \chi_{ia}\rangle = \langle \hat{q}_\alpha \rangle = 0$) vanish. Consequently, all uncertainties reduce to $\Delta q_{\alpha}= \sqrt{\langle \hat{q}^2_{\alpha} \rangle}$ and $\Delta p_{\alpha}= \sqrt{\langle \hat{p}^2_{\alpha} \rangle}$.  These uncertainties encode the complete Gaussian statistics of the dressed photon mode $\alpha$, and therefore provide a direct handle on how light-matter coupling and many-body electronic correlations modify the quantum fluctuations of the electromagnetic field.  
In particular, deviations of $(\Delta q_{\alpha}, \Delta p_{\alpha})$ from their vacuum values quantify the degree of squeezing induced by collective electronic response and photon hybridization.  
To quantify this squeezing, we calculate the squeezing parameter $r_{\alpha}$ of mode $\alpha$, which can be written as~\cite{Kam2024SuperPoissonian}
\begin{align}
r_{\alpha}
= \frac{1}{2}
\ln\!\left(
\frac{\Delta q_{\alpha}}{\Delta p_{\alpha}/\omega_{\alpha}}
\right)
=
\frac{1}{4}
\ln\!\left(
\frac{\langle \hat{q}_{\alpha}^{2} \rangle \, \omega_{\alpha}}
{\langle \hat{p}_{\alpha}^{2} \rangle / \omega_{\alpha}}
\right).
\label{eq:r}
\end{align}
This formula makes it explicit that the effective squeezing arises from the 
renormalized quadrature variances $\langle \hat q_{\alpha}^{2} \rangle$ and $\langle \hat p_{\alpha}^{2} \rangle$ generated by the Bogoliubov 
coefficients $(X,Y)$.
Importantly, this squeezing is not imposed externally but emerges from higher-order polarization effects and the
light-matter dressing of the ground state, providing a direct link
between electronic correlations and nonclassical photon statistics.
The corresponding photon number is then given by the following equation
\begin{align}
\langle \hat n_{\alpha} \rangle
= \frac{\omega_{\alpha} \langle \hat{q}_{\alpha}^{2} \rangle
      + \frac{1}{\omega_{\alpha}} \langle \hat{p}_{\alpha}^{2} \rangle - 1}{2}
= \big( Y^{\mathrm T} Y \big)_{\alpha,\alpha},
\label{eq:photon-number}
\end{align}
which quantifies the virtual photon population generated by strong light-matter coupling. The photon number has previously been analyzed within optimized-effective-potential (OEP) approaches \cite{pellegrini2015}  and can also be estimated using the photon-GA functional \cite{flick2022}. However, these earlier exchange-correlation approximations incorporate only first-order photon-mediated corrections and therefore can be expected to break down when higher-order light-matter correlations or multi-photon processes become important. In contrast, the pMBD functional includes the full hierarchy of higher-order mixed electron-photon contributions, enabling us to systematically assess how these effects modify the exchange-correlation energy, virtual photon populations, and nonclassical photon statistics in strongly coupled systems. 
Furthermore, the second-order photon correlation function $\big\langle 
\hat a_{\alpha}^\dagger 
\hat a_{\alpha}^\dagger 
\hat a_{\alpha} 
\hat a_{\alpha} 
\big\rangle$, which is central for 
quantifying photon statistics and the Mandel $Q$ parameter~\cite{mandel1979fluorescence}, 
is defined as
\begin{align*}
\big\langle 
\hat a_{\alpha}^\dagger 
\hat a_{\alpha}^\dagger 
\hat a_{\alpha} 
\hat a_{\alpha} 
\big\rangle
=
2\big( Y^{\mathrm T} Y \big)_{\alpha,\alpha}^{2}
+
\big( X^{\mathrm T} Y \big)_{\alpha,\alpha}^{2}.
\end{align*}
From these quantities, the Mandel $Q$ parameter follows directly as
\begin{align}
Q_\alpha
=
\frac{
\big( Y^{\mathrm T} Y \big)_{\alpha,\alpha}^{2}
+
\big( X^{\mathrm T} Y \big)_{\alpha,\alpha}^{2}
}{
\big( Y^{\mathrm T} Y \big)_{\alpha,\alpha}
},
\label{eq:Q}
\end{align}
highlighting how photon-number fluctuations and higher-order light-matter correlations are encoded in the Bogoliubov coefficients. The Mandel $Q$ parameter provides a simple way to distinguish photon statistics~\cite{mandel1995optical}: $Q<0$ indicates sub-Poissonian (antibunched) light, $Q=0$ corresponds to Poissonian (coherent) light, and $Q>0$ reflects super-Poissonian behavior typically found in thermal or chaotic classical sources. 

Finally, to quantify the entanglement between the cavity mode and the collective electronic excitations, we compute the Von Neumann entropy of the photonic subsystem, $S_{vN} = -\mathrm{Tr}(\hat{\rho}_{\alpha} \ln \hat{\rho}_{\alpha})$. Since the pMBD ground state is a Gaussian state, the reduced density matrix of the cavity mode $\hat{\rho}_{\alpha}$ is fully characterized by its covariance matrix~\cite{weedbrook2012,Serafini2017}.
The entropy is determined solely by the symplectic eigenvalue $\nu_\alpha$ of this covariance matrix, which, given the absence of $x$-$p$ correlations in the diagonalized frame, is defined by the uncertainty product $\nu_\alpha = 2 \sqrt{\langle \Delta \hat{q}_\alpha^2 \rangle \langle \Delta \hat{p}_\alpha^2 \rangle}$. The Von Neumann entropy is then given by~\cite{gori2023second}
\begin{align}
    S_{vN} = \frac{\nu_\alpha + 1}{2} \ln \left( \frac{\nu_\alpha + 1}{2} \right) - \frac{\nu_\alpha - 1}{2} \ln \left( \frac{\nu_\alpha - 1}{2} \right).
    \label{eq:S}
\end{align}
For a pure, non-entangled state (such as the vacuum state), the Heisenberg limit is saturated ($\nu_\alpha=1$), yielding $S_{vN}=0$.
Values of $\nu_\alpha > 1$ (and thus $S_{vN} > 0$) indicate that the cavity mode is in a mixed state due to intrinsic entanglement with the electronic degrees of freedom.

\textit{Application:} In the following, we apply the developed framework to one-dimensional chains of Ar atoms with a varying number of Ar atoms. These atoms are aligned in the z-axis, collectively coupled to a single cavity mode with polarization in the z-direction. For all results shown in this work, we apply a photon frequency of $\omega_\alpha=2$~eV, a distance between Ar atoms of $d=4$\AA ~and a light-matter coupling strength of $\lambda_\alpha=0.025$ a.u. This geometry allows us to isolate how collective coupling modifies the ground-state energy, virtual photon population, and non-classical photon statistics as the system size increases.
\begin{figure}[t]
  \centering
  \includegraphics[width=0.5\textwidth]{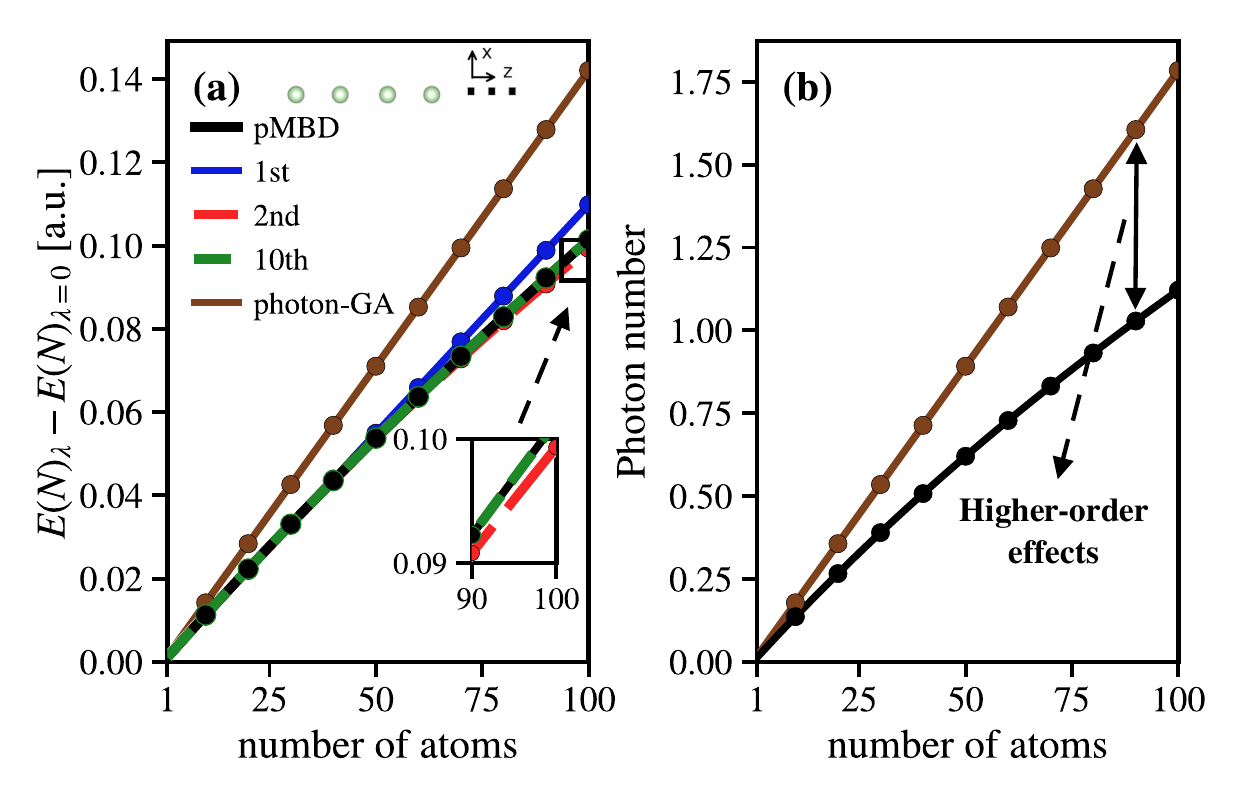}
\caption{
\textbf{Electron-photon exchange-correlation energy and photon number for chains of Ar atoms.} (a) Electron-photon exchange-correlation energy predicted by pMBD (black), photon-GA (brown), and pMBD up to first (blue), second (red), and tenth order (green). The inset shows deviations between the second order and tenth/full pMBD results for a large number of atoms (90-100). In (b), we compare the ground-state photon number of pMBD (black) and photon-GA (brown).
}
\label{fig:1}
\end{figure}

Fig.~\ref{fig:1} illustrates the scaling behavior of the electron-photon interaction energy and the virtual photon population $\langle \hat n_\alpha\rangle$ (Eq.~\ref{eq:photon-number}) as a function of system size for a linear chain of Ar atoms in an optical cavity as shown schematically in the figure. Specifically, Fig.~1(a) plots the difference in energy between coupled and uncoupled, $E(N)_\lambda - E(N)_{\lambda=0}$, as a function of the number of atoms $N$. The data compares the full pMBD result (black dash-dotted line) with the photon-GA functional (brown line with circles) and various orders of the pMBD many-body expansion: 1st order (blue solid line), 2nd order (red solid line), and 10th order (green dashed line) using the order-by-order expansion via frequency-integration. 
The inset provides a magnified view of the large-$N$ limit ($N=90$ to $100$).
From Fig.~\ref{fig:1}(a), we see that both the first-order approximation and the photon-GA functional scale linearly with the number of atoms $N$~\cite{flick2018a}. In contrast, the second and higher orders progressively converge toward the full pMBD result, which exhibits a sublinear growth. Using the pMBD functional, we find that, as $N$ increases, higher-order processes lead to a deviation of accumulating energy additively~\cite{novokreschenov2023}, resulting in a reduced effective contribution per atom that simple single-photon functionals, such as the photon GA functional, cannot capture. Fig.~\ref{fig:1}(b) displays the corresponding total photon number as a function of the number of atoms $N$, explicitly contrasting the photon-GA prediction (brown) with the pMBD result (black). While the photon-GA functional predicts a linear increase of $\langle \hat{n}_\alpha \rangle$ with system size again, the pMBD result grows more slowly and displays a clear sublinear trend. 
This difference further highlights the collective nature of the interaction: as more atoms couple to the cavity, higher-order effects redistribute the virtual photon population, deviating from the linear prediction of independent-particle models.

\begin{figure}[t] 
\centering 
\includegraphics[width=1.0\linewidth]{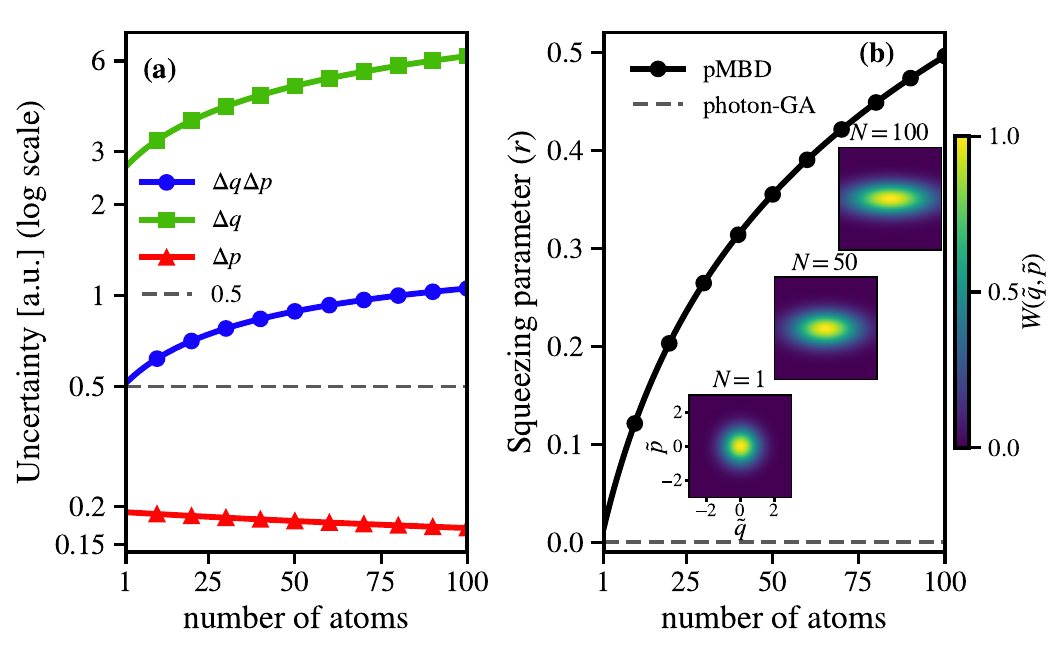}
\caption{\textbf{Uncertainties and squeezing parameters for chains of Ar atoms.} (a) Cavity quadrature uncertainties $\Delta q$ (green squares) and $\Delta p$ (red triangles), and their uncertainty product $\Delta q \Delta p$ (blue circles), as a function of the number of atoms $N$. The gray dashed line marks the Heisenberg uncertainty limit ($0.5$). (b) The squeezing parameter $r$ predicted by pMBD (black circles), whereas the photon-GA functional (gray dashed line) predicts no squeezing ($r=0$). Insets display the Wigner quasiprobability distributions $W(\tilde{q}, \tilde{p})$ in dimensionless phase-space coordinates ($\tilde{q}=\omega_{\alpha}q, \tilde{p}=p/\omega_{\alpha}$) for $N=1, 50,$ and $100$, visualizing the phase-space deformation predicted by pMBD.}
\label{fig:2}
\end{figure}

Fig.~\ref{fig:2} characterizes the quantum statistical properties of the cavity mode as a function of the number of atoms $N$. 
Fig.~\ref{fig:2}(a) plots the quadrature uncertainties predicted by pMBD: the position uncertainty $\Delta q_\alpha$ (green squares), the momentum uncertainty $\Delta p_\alpha$ (red triangles), and their Heisenberg product $\Delta q_\alpha \Delta p_\alpha$ (blue circles). 
The theoretical vacuum limit of $0.5$~\cite{Walls1983} is indicated by the grey dashed line for reference.
From Fig.~\ref{fig:2}(a), we see how collective light-matter interactions reshape the quantum fluctuations of the cavity mode. 
As the number of atoms $N$ increases, the quadrature uncertainties diverge from the vacuum level: the state becomes increasingly squeezed in the $p$-quadrature while expanding in the $q$-quadrature. Similar features were observed in Ref.~\cite{tang2025deep} for an increase in $\lambda_\alpha$ instead of the number of atoms N. Fig.~\ref{fig:2}(b) displays the squeezing parameter $r$ (Eq.~\ref{eq:r}) calculated via the pMBD method (red circles) compared to the photon-GA prediction (grey dashed line), which remains zero. 
The insets in Fig.~\ref{fig:2}(b) visualize the corresponding Wigner distributions of the cavity photon state in the pMBD framework at three distinct system sizes ($N=1, 50,$ and $100$), illustrating the deformation of the state's fluctuations in phase space. This phase-space deformation is highlighted by the Wigner distributions, which stretch from a vacuum circle into an elongated ellipse as $N$ increases. 
This visualizes the quantitative trend of the squeezing parameter $r$, confirming that the reduction in momentum uncertainty corresponds to a coherent squeezing of the vacuum state, despite the overall increase in the uncertainty product.

Finally, Fig.~\ref{fig:3} characterizes the quantum nature of the cavity fluctuations by analyzing the photon statistics. 
Fig.~\ref{fig:3}(a) displays the Mandel $Q$  parameter (Eq.~\ref{eq:Q}) (blue squares) and the photon number variance $\Delta n_\alpha^2$ (red circles) as a function of the number of atoms $N$. 
The results from the pMBD method (solid lines) are contrasted with the photon-GA approximation (dashed lines). The dotted grey line at $Q=0$ represents the Poissonian limit characteristic of coherent states; values below this line ($Q<0$) indicate sub-Poissonian statistics, while values above ($Q>0$) indicate super-Poissonian behavior. Here, pMBD predicts strongly super-Poissonian statistics with $Q>2$ for $N=100$ with significant values as well in the photon number variance $\Delta n_\alpha^2$. In contrast, the photon-GA method (dashed lines) predicts sub-Poissonian statistics ($Q<0$). However, this is not a physical signature of quantum antibunching, but rather a failure of the approximation. Since the photon-GA method includes only single-photon processes, the two-photon correlation function vanishes ($\langle \hat a^\dagger_\alpha \hat a^\dagger_\alpha \hat a_\alpha \hat a_\alpha \rangle = 0$). 
Consequently, the photon number variance is artificially truncated to $\Delta n_\alpha^2 = \langle \hat n_\alpha^2 \rangle - \langle \hat n_\alpha \rangle^2$, forcing the Mandel parameter to be negative ($Q = -\langle \hat n_\alpha \rangle$). 
Thus, the observed sub-Poissonian statistics arise from an incorrect truncation of the two-photon Hilbert space, rather than a genuine quantum origin. 
Fig.~\ref{fig:3}(b) investigates the nature of the fluctuations by plotting the von Neumann entropy $S_{\mathrm{vN}}$ (Eq.~\ref{eq:S}) against the mean photon number. 
The pMBD results (black diamonds) are compared to the photon-GA predictions (red circles) and the theoretical thermal limit (black dashed line), which represents the maximum entropy for a given photon number. 
In contrast to photon-GA in Fig.\ref{fig:3}~(a), pMBD predicts a physical transition to super-Poissonian statistics ($Q > 0$). 
While such broadened distributions are often associated with signatures of classical thermal noise, Fig.~\ref{fig:3}(b) rules out thermalization as the cause. 
The pMBD entropy reveals a significant deviation from a thermal distribution, as the entropy values are significantly below the thermal limit. 
This gap demonstrates that the super-Poissonian statistics observed in Fig.~\ref{fig:3}(a) are not driven by thermal mixing, but rather by the squeezed vacuum nature of the state. 
As the phase space ellipse stretches (see Fig.~\ref{fig:2}), it intersects higher photon number states, creating large number fluctuations ($\Delta n_\alpha^2 > \langle \hat n_\alpha \rangle$) while maintaining the low entropy of a pure quantum state. 
This observation of super-Poissonian squeezed light aligns with recent analytical solutions of the deep-strong-coupling Quantum Rabi Model~\cite{Kam2024SuperPoissonian}. We also note that a squeezed wave function ansatz has been recently successfully used for QED Hartree-Fock~\cite{Mazin2024}.
\begin{figure}[t]
  \centering
  \includegraphics[width=1.0\linewidth]{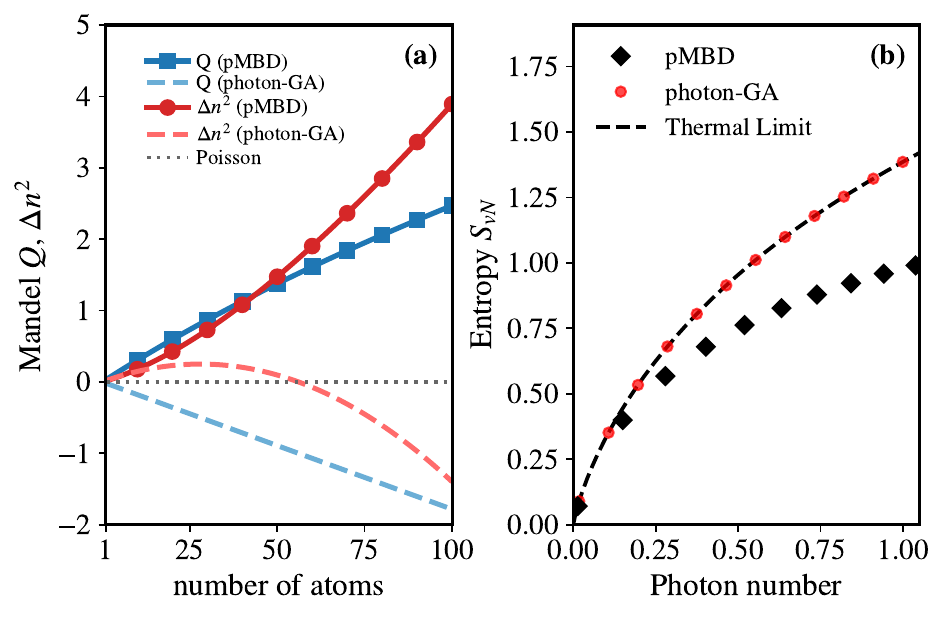}
  \caption{\textbf{Entanglement and correlations for chains of Ar atoms.} (a) Mandel $Q$ parameter (blue squares) and photon number variance $\Delta n^2$ (red circles). The gray dotted line marks the Poissonian limit ($Q=0$).  (b) Von Neumann entanglement entropy $S_{vN}$ vs photon number. The pMBD (black diamonds) predicts less entropy than the thermal limit (dashed black line), whereas the photon-GA (red circles) predicts a thermal-like behavior.
  }
  \label{fig:3}
\end{figure}

\textit{Conclusion:} In summary, we have introduced a framework for computing photonic observables in strongly coupled light–matter systems within the pMBD framework of QEDFT. This approach captures collective, higher-order, and multi-photon processes that emerge when many emitters interact with a single cavity mode. 
Applications to chains of Ar atoms demonstrate that collective coupling leads to sublinear photon number scaling and the emergence of non-classical photon statistics. These results show that many-body and multi-photon contributions are essential for a consistent description of both energetic and photonic properties, beyond the scope of simpler functionals that truncate the photonic subspace. 

\textit{Acknowledgments:} We acknowledge funding from the Defense Advanced Research Projects Agency (DARPA), and startup funding from the City College of New York. 
The Flatiron Institute is a division of the Simons Foundation.

\bibliography{refs} 

@article{mejia2025meta,
  title={A Meta-Generalized Gradient Approximation for the Cavity-Dependent Exchange--Correlation Interaction in Strongly Coupled Light--Matter Systems},
  author={Mejia-Rodriguez, Daniel and Govind, Niranjan},
  journal={The Journal of Physical Chemistry Letters},
  volume={16},
  pages={13139--13146},
  year={2025},
  publisher={American Chemical Society},
  doi={10.1021/acs.jpclett.5c02285},
  url={https://doi.org/10.1021/acs.jpclett.5c02285}
}

@article{Mazin2024,
  title = {Light-Matter Hybridization and Entanglement from First-Principles},
  author = {Mazin, Ilia M. and Zhang, Yu},
  journal = {arXiv preprint arXiv:2411.15022},
  year = {2024},
  url = {https://arxiv.org/abs/2411.15022}
}

@article{Carmichael1985,
  title = {Photon Antibunching and Squeezing for a Single Atom in a Resonant Cavity},
  author = {Carmichael, H. J.},
  journal = {Phys. Rev. Lett.},
  volume = {55},
  number = {25},
  pages = {2790--2793},
  year = {1985},
  publisher = {American Physical Society},
  doi = {10.1103/PhysRevLett.55.2790}
}

@article{Rempe1991,
  title = {Optical bistability and photon statistics in cavity quantum electrodynamics},
  author = {Rempe, G. and Thompson, R. J. and Brecha, R. J. and Lee, W. D. and Kimble, H. J.},
  journal = {Phys. Rev. Lett.},
  volume = {67},
  number = {13},
  pages = {1727--1730},
  year = {1991},
  publisher = {American Physical Society},
  doi = {10.1103/PhysRevLett.67.1727}
}

@article{Walls1983,
  title = {Squeezed states of light},
  author = {Walls, D. F.},
  journal = {Nature},
  volume = {306},
  number = {5939},
  pages = {141--146},
  year = {1983},
  publisher = {Nature Publishing Group},
  doi = {10.1038/306141a0}
}

@article{DeVoe1996,
  title = {Observation of Superradiant and Subradiant Spontaneous Emission of Two Trapped Ions},
  author = {DeVoe, R. G. and Brewer, R. G.},
  journal = {Phys. Rev. Lett.},
  volume = {76},
  number = {12},
  pages = {2049--2052},
  year = {1996},
  publisher = {American Physical Society},
  doi = {10.1103/PhysRevLett.76.2049}
}

@article{Brandes2005,
  title = {Coherent and collective quantum optical effects in mesoscopic systems},
  author = {Brandes, Tobias},
  journal = {Phys. Rep.},
  volume = {408},
  number = {3-4},
  pages = {315--407},
  year = {2005},
  publisher = {Elsevier},
  doi = {10.1016/j.physrep.2004.12.002}
}

@article{Basche1992,
  title = {Photon antibunching in the fluorescence of a single dye molecule trapped in a solid},
  author = {Basch\'e, T. and Moerner, W. E. and Orrit, M. and Tallet, H.},
  journal = {Phys. Rev. Lett.},
  volume = {69},
  number = {10},
  pages = {1516--1519},
  year = {1992},
  publisher = {American Physical Society},
  doi = {10.1103/PhysRevLett.69.1516}
}

@article{Michler2000,
  title = {A Quantum Dot Single-Photon Turnstile Device},
  author = {Michler, P. and Kiraz, A. and Becher, C. and Schoenfeld, W. V. and Petroff, P. M. and Zhang, L. and Hu, E. and Imamoglu, A.},
  journal = {Science},
  volume = {290},
  number = {5500},
  pages = {2282--2285},
  year = {2000},
  publisher = {American Association for the Advancement of Science},
  doi = {10.1126/science.290.5500.2282}
}

@article{Breitenbach1997,
  title = {Measurement of the quantum states of squeezed light},
  author = {Breitenbach, G. and Schiller, S. and Mlynek, J.},
  journal = {Nature},
  volume = {387},
  pages = {471--475},
  year = {1997},
  publisher = {Nature Publishing Group},
  doi = {10.1038/387471a0}
}

@article{Polzik1992,
  title = {Spectroscopy with squeezed light},
  author = {Polzik, E. S. and Carri, J. and Kimble, H. J.},
  journal = {Phys. Rev. Lett.},
  volume = {68},
  number = {20},
  pages = {3020--3023},
  year = {1992},
  publisher = {American Physical Society},
  doi = {10.1103/PhysRevLett.68.3020}
}

@book{Serafini2017,
  title={Quantum Continuous Variables: A Primer of Theoretical Methods},
  author={Serafini, Alessio},
  year={2017},
  publisher={CRC Press},
  address={Boca Raton},
  isbn={9781482246346}
}

@article{Kam2024SuperPoissonian,
  author        = {Kam, Chon-Fai and Hu, Xuedong},
  title         = {Super-Poissonian Squeezed Light in the Deep Strong Regime of the Quantum Rabi Model},
  journal       = {arXiv preprint},
  archivePrefix = {arXiv},
  primaryClass  = {quant-ph},
  year          = {2024},
  doi           = {10.48550/arXiv.2412.04085}
}

@article{mandel1979fluorescence,
author = {L. Mandel},
journal = {Opt. Lett.},
number = {7},
pages = {205--207},
publisher = {Optica Publishing Group},
title = {Sub-Poissonian photon statistics in resonance fluorescence},
volume = {4},
month = {Jul},
year = {1979},
url = {https://opg.optica.org/ol/abstract.cfm?URI=ol-4-7-205},
doi = {10.1364/OL.4.000205},
}

@article{schaefer2020relevance,
author = {Schäfer, Christian and Ruggenthaler, Michael and Rokaj, Vasil and Rubio, Angel},
title = {Relevance of the Quadratic Diamagnetic and Self-Polarization Terms in Cavity Quantum Electrodynamics},
journal = {ACS Photonics},
volume = {7},
number = {4},
pages = {975-990},
year = {2020},
doi = {10.1021/acsphotonics.9b01649},
note ={PMID: 32322607},
URL = {https://doi.org/10.1021/acsphotonics.9b01649},
eprint = {https://doi.org/10.1021/acsphotonics.9b01649}
}

@article{flick2022,
  title = {Simple Exchange-Correlation Energy Functionals for Strongly Coupled Light-Matter Systems Based on the Fluctuation-Dissipation Theorem},
  author = {Flick, Johannes},
  journal = {Phys. Rev. Lett.},
  volume = {129},
  issue = {14},
  pages = {143201},
  numpages = {7},
  year = {2022},
  month = {Sep},
  publisher = {American Physical Society},
  doi = {10.1103/PhysRevLett.129.143201},
  url = {https://link.aps.org/doi/10.1103/PhysRevLett.129.143201}
}

@article{flick2018a,
author = {Flick, Johannes and Schäfer, Christian and Ruggenthaler, Michael and Appel, Heiko and Rubio, Angel},
title = {Ab Initio Optimized Effective Potentials for Real Molecules in Optical Cavities: Photon Contributions to the Molecular Ground State},
journal = {ACS Photonics},
volume = {5},
number = {3},
pages = {992-1005},
year = {2018},
doi = {10.1021/acsphotonics.7b01279}
}

@article{gori2023second,
  title={Second quantization of many-body dispersion interactions for chemical and biological systems},
  author={Gori, Matteo and Kurian, Philip and Tkatchenko, Alexandre},
  journal={Nature Communications},
  volume={14},
  number={1},
  pages={8218},
  year={2023},
  publisher={Nature Publishing Group},
  doi={10.1038/s41467-023-43785-z},
  url={https://doi.org/10.1038/s41467-023-43785-z}
}

@article{tasci2025pmbd,
  title        = {Photon Many‐Body Dispersion: An Exchange–Correlation Functional for Strongly Coupled Light–Matter Systems},
  author       = {Tasci, Cankut and Cunha, Leonardo A. and Flick, Johannes},
  journal      = {Physical Review Letters},
  volume       = {134},
  number       = {7},
  pages        = {073002},
  year         = {2025},
  publisher    = {American Physical Society},
  doi          = {10.1103/PhysRevLett.134.073002}
}

@book{mandel1995optical,
  title     = {Optical Coherence and Quantum Optics},
  author    = {Mandel, Leonard and Wolf, Emil},
  year      = {1995},
  publisher = {Cambridge University Press}
}

@article{vidal2021manipulating,
author = {Francisco J. Garcia-Vidal  and Cristiano Ciuti  and Thomas W. Ebbesen },
title = {Manipulating matter by strong coupling to vacuum fields},
journal = {Science},
volume = {373},
number = {6551},
pages = {eabd0336},
year = {2021},
doi = {10.1126/science.abd0336},
URL = {https://www.science.org/doi/abs/10.1126/science.abd0336},
eprint = {https://www.science.org/doi/pdf/10.1126/science.abd0336}}

@article{ruggenthaler2014,
  title = {Quantum-electrodynamical density-functional theory: Bridging quantum optics and electronic-structure theory},
  author = {Ruggenthaler, Michael and Flick, Johannes and Pellegrini, Camilla and Appel, Heiko and Tokatly, Ilya V. and Rubio, Angel},
  journal = {Phys. Rev. A},
  volume = {90},
  issue = {1},
  pages = {012508},
  numpages = {26},
  year = {2014},
  month = {Jul},
  publisher = {American Physical Society},
  doi = {10.1103/PhysRevA.90.012508},
  url = {https://link.aps.org/doi/10.1103/PhysRevA.90.012508}
}

@article{tokatly2013,
  title = {Time-Dependent Density Functional Theory for Many-Electron Systems Interacting with Cavity Photons},
  author = {Tokatly, I. V.},
  journal = {Phys. Rev. Lett.},
  volume = {110},
  issue = {23},
  pages = {233001},
  numpages = {5},
  year = {2013},
  month = {Jun},
  publisher = {American Physical Society},
  doi = {10.1103/PhysRevLett.110.233001},
  url = {https://link.aps.org/doi/10.1103/PhysRevLett.110.233001}
}

@article{tang2025deep,
    author = {Tang, Yifan and Andolina, Gian Marcello and Cuzzocrea, Alice and Mezera, Matěj and Szabó, P. Bernát and Schätzle, Zeno and Noé, Frank and Erdman, Paolo A.},
    title = {Deep quantum Monte Carlo approach for polaritonic chemistry},
    journal = {The Journal of Chemical Physics},
    volume = {163},
    number = {3},
    pages = {034108},
    year = {2025},
    month = {07},
    issn = {0021-9606},
    doi = {10.1063/5.0272805},
    url = {https://doi.org/10.1063/5.0272805},
    eprint = {https://pubs.aip.org/aip/jcp/article-pdf/doi/10.1063/5.0272805/20590434/034108_1_5.0272805.pdf},
}

@article{novokreschenov2023,
  title = {Quantum electrodynamical density functional theory for generalized Dicke model},
  author = {Novokreschenov, D. and Kudlis, A. and Iorsh, I. and Tokatly, I. V.},
  journal = {Phys. Rev. B},
  volume = {108},
  issue = {23},
  pages = {235424},
  numpages = {18},
  year = {2023},
  month = {Dec},
  publisher = {American Physical Society},
  doi = {10.1103/PhysRevB.108.235424},
  url = {https://link.aps.org/doi/10.1103/PhysRevB.108.235424}
}

@article{pellegrini2015,
  title = {Optimized Effective Potential for Quantum Electrodynamical Time-Dependent Density Functional Theory},
  author = {Pellegrini, Camilla and Flick, Johannes and Tokatly, Ilya V. and Appel, Heiko and Rubio, Angel},
  journal = {Phys. Rev. Lett.},
  volume = {115},
  issue = {9},
  pages = {093001},
  numpages = {5},
  year = {2015},
  month = {Aug},
  publisher = {American Physical Society},
  doi = {10.1103/PhysRevLett.115.093001},
  url = {https://link.aps.org/doi/10.1103/PhysRevLett.115.093001}
}

@article{Cao_2025, title={Cavity-Induced Quantum Interference and Collective Interactions in van der Waals Systems}, volume={16}, ISSN={1948-7185}, url={http://dx.doi.org/10.1021/acs.jpclett.5c00031}, DOI={10.1021/acs.jpclett.5c00031}, number={22}, journal={The Journal of Physical Chemistry Letters}, publisher={American Chemical Society (ACS)}, author={Cao, Jianshu and Pollak, Eli}, year={2025}, month=may, pages={5466–5472} }

@article{COLPA1978327,
title = {Diagonalization of the quadratic boson hamiltonian},
journal = {Physica A: Statistical Mechanics and its Applications},
volume = {93},
number = {3},
pages = {327-353},
year = {1978},
issn = {0378-4371},
doi = {https://doi.org/10.1016/0378-4371(78)90160-7},
url = {https://www.sciencedirect.com/science/article/pii/0378437178901607},
author = {J.H.P. Colpa}
}

@article{weedbrook2012,
  title = {Gaussian quantum information},
  author = {Weedbrook, Christian and Pirandola, Stefano and Garc\'{\i}a-Patr\'on, Ra\'ul and Cerf, Nicolas J. and Ralph, Timothy C. and Shapiro, Jeffrey H. and Lloyd, Seth},
  journal = {Rev. Mod. Phys.},
  volume = {84},
  issue = {2},
  pages = {621--669},
  numpages = {0},
  year = {2012},
  month = {May},
  publisher = {American Physical Society},
  doi = {10.1103/RevModPhys.84.621},
  url = {https://link.aps.org/doi/10.1103/RevModPhys.84.621}
}

@article{tkatchenko2009,
  title = {Accurate Molecular Van Der Waals Interactions from Ground-State Electron Density and Free-Atom Reference Data},
  author = {Tkatchenko, Alexandre and Scheffler, Matthias},
  journal = {Phys. Rev. Lett.},
  volume = {102},
  issue = {7},
  pages = {073005},
  numpages = {4},
  year = {2009},
  month = {Feb},
  publisher = {American Physical Society},
  doi = {10.1103/PhysRevLett.102.073005},
  url = {https://link.aps.org/doi/10.1103/PhysRevLett.102.073005}
}

@article{donchev2006,
    author = {Donchev, A. G.},
    title = "{Many-body effects of dispersion interaction}",
    journal = {The Journal of Chemical Physics},
    volume = {125},
    number = {7},
    pages = {074713},
    year = {2006},
    month = {08},
    issn = {0021-9606},
    doi = {10.1063/1.2337283},
    url = {https://doi.org/10.1063/1.2337283},
    eprint = {https://pubs.aip.org/aip/jcp/article-pdf/doi/10.1063/1.2337283/15390550/074713\_1\_online.pdf},
}

@article{tkatchenko2012,
  title = {Accurate and Efficient Method for Many-Body van der Waals Interactions},
  author = {Tkatchenko, Alexandre and DiStasio, Robert A. and Car, Roberto and Scheffler, Matthias},
  journal = {Phys. Rev. Lett.},
  volume = {108},
  issue = {23},
  pages = {236402},
  numpages = {5},
  year = {2012},
  month = {Jun},
  publisher = {American Physical Society},
  doi = {10.1103/PhysRevLett.108.236402},
  url = {https://link.aps.org/doi/10.1103/PhysRevLett.108.236402}
}

@article{tkatchenko2013,
    author = {Tkatchenko, Alexandre and Ambrosetti, Alberto and DiStasio, Robert A., Jr.},
    title = "{Interatomic methods for the dispersion energy derived from the adiabatic connection fluctuation-dissipation theorem}",
    journal = {The Journal of Chemical Physics},
    volume = {138},
    number = {7},
    pages = {074106},
    year = {2013},
    month = {02},
    issn = {0021-9606},
    doi = {10.1063/1.4789814},
    url = {https://doi.org/10.1063/1.4789814},
    eprint = {https://pubs.aip.org/aip/jcp/article-pdf/doi/10.1063/1.4789814/14045058/074106\_1\_online.pdf},
}

@article{patrahau2024,
author = {Patrahau, B. and Piejko, M. and Mayer, R. J. and Antheaume, C. and Sangchai, T. and Ragazzon, G. and Jayachandran, A. and Devaux, E. and Genet, C. and Moran, J. and Ebbesen, T. W.},
title = {Direct Observation of Polaritonic Chemistry by Nuclear Magnetic Resonance Spectroscopy},
journal = {Angewandte Chemie International Edition},
volume = {63},
number = {23},
pages = {e202401368},
doi = {https://doi.org/10.1002/anie.202401368},
url = {https://onlinelibrary.wiley.com/doi/abs/10.1002/anie.202401368},
eprint = {https://onlinelibrary.wiley.com/doi/pdf/10.1002/anie.202401368},
year = {2024}
}

@article{haugland2023understanding,
    author = {Haugland, Tor S. and Philbin, John P. and Ghosh, Tushar K. and Chen, Ming and Koch, Henrik and Narang, Prineha},
    title = {Understanding the polaritonic ground state in cavity quantum electrodynamics},
    journal = {The Journal of Chemical Physics},
    volume = {162},
    number = {19},
    pages = {194106},
    year = {2025},
    month = {05},
    issn = {0021-9606},
    doi = {10.1063/5.0258935},
    url = {https://doi.org/10.1063/5.0258935},
    eprint = {https://pubs.aip.org/aip/jcp/article-pdf/doi/10.1063/5.0258935/20522437/194106\_1\_5.0258935.pdf},
}

@article{philbin2023molecular,
author = {Philbin, John P. and Haugland, Tor S. and Ghosh, Tushar K. and Ronca, Enrico and Chen, Ming and Narang, Prineha and Koch, Henrik},
title = {Molecular van der Waals Fluids in Cavity Quantum Electrodynamics},
journal = {The Journal of Physical Chemistry Letters},
volume = {14},
number = {40},
pages = {8988-8993},
year = {2023},
doi = {10.1021/acs.jpclett.3c01790},
URL = {https://doi.org/10.1021/acs.jpclett.3c01790},
eprint = {https://doi.org/10.1021/acs.jpclett.3c01790}
}

@article{ahn2023modification,
author = {Wonmi Ahn  and Johan F. Triana  and Felipe Recabal  and Felipe Herrera  and Blake S. Simpkins },
title = {Modification of ground-state chemical reactivity via light–matter coherence in infrared cavities},
journal = {Science},
volume = {380},
number = {6650},
pages = {1165-1168},
year = {2023},
doi = {10.1126/science.ade7147},
URL = {https://www.science.org/doi/abs/10.1126/science.ade7147},
eprint = {https://www.science.org/doi/pdf/10.1126/science.ade7147}}

@article{thomas2019tilting,
author = {A. Thomas  and L. Lethuillier-Karl  and K. Nagarajan  and R. M. A. Vergauwe  and J. George  and T. Chervy  and A. Shalabney  and E. Devaux  and C. Genet  and J. Moran  and T. W. Ebbesen },
title = {Tilting a ground-state reactivity landscape by vibrational strong coupling},
journal = {Science},
volume = {363},
number = {6427},
pages = {615-619},
year = {2019},
doi = {10.1126/science.aau7742},
URL = {https://www.science.org/doi/abs/10.1126/science.aau7742},
eprint = {https://www.science.org/doi/pdf/10.1126/science.aau7742}}

@article{bxiang2020i,
author = {Bo Xiang  and Raphael F. Ribeiro  and Matthew Du  and Liying Chen  and Zimo Yang  and Jiaxi Wang  and Joel Yuen-Zhou  and Wei Xiong },
title = {Intermolecular vibrational energy transfer enabled by microcavity strong light–matter coupling},
journal = {Science},
volume = {368},
number = {6491},
pages = {665-667},
year = {2020},
doi = {10.1126/science.aba3544},
URL = {https://www.science.org/doi/abs/10.1126/science.aba3544},
eprint = {https://www.science.org/doi/pdf/10.1126/science.aba3544}}

@article{ambrosetti2014long,
    author = {Ambrosetti, Alberto and Reilly, Anthony M. and DiStasio, Robert A., Jr. and Tkatchenko, Alexandre},
    title = "{Long-range correlation energy calculated from coupled atomic response functions}",
    journal = {The Journal of Chemical Physics},
    volume = {140},
    number = {18},
    pages = {18A508},
    year = {2014},
    month = {02},
    issn = {0021-9606},
    doi = {10.1063/1.4865104},
    url = {https://doi.org/10.1063/1.4865104},  
}

@article{lu2024electronphoton,
  title = {Electron-photon exchange-correlation approximation for quantum-electrodynamical density-functional theory},
  author = {Lu, I-Te and Ruggenthaler, Michael and Tancogne-Dejean, Nicolas and Latini, Simone and Penz, Markus and Rubio, Angel},
  journal = {Physical Review A},
  volume = {109},
  number = {5},
  pages = {052823},
  year = {2024},
  month = {May},
  doi = {10.1103/PhysRevA.109.052823},
  url = {https://link.aps.org/doi/10.1103/PhysRevA.109.052823},
  publisher = {American Physical Society},
  numpages = {15}
}

@article{schafer2021making,
  title={Making Ab Initio QED Functional(s): Nonperturbative and Photon-Free Effective Frameworks for Strong Light-Matter Coupling},
  author={Sch{\"a}fer, C. and Buchholz, F. and Penz, M. and Ruggenthaler, M. and Rubio, A.},
  journal={Proc. Natl. Acad. Sci. U.S.A.},
  volume={118},
  pages={e2110464118},
  year={2021},
  doi={10.1073/pnas.2110464118}
}

\end{document}